# Self-induced transparency mode-locking in a Ti:sapphire laser with an intracavity rubidium cell

M.V. Arkhipov[1,2], R.M. Arkhipov[1,2,3], A.A. Shimko[1], I. Babushkin[4,5,6],

N.N. Rosanov[2,3,7]

[1] St. Petersburg State University, 7/9 Universitetskaya nab., St. Petersburg, 199034, Russia
[2] ITMO University, Kronverkskiy Prospekt 49, St. Petersburg, 197101, Russia
[3] Ioffe Institute, Politekhnicheskaya str. 26, St. Petersburg, 194021, Russia

[4] Institute of Quantum Optics, Welfengarten 1, 30167, Hannover, Germany

[5] Cluster of Excellence PhoenixD (Photonics, Optics, and Engineering – Innovation Across Disciplines), Hannover, Germany

[6] Max Born Institute, Max-Born-Strasse 2a, Berlin, 10117, Germany

[7] Vavilov State Optical Institute, Kadetskaya Liniya v.o. 5/2, St. Petersburg, 199053, Russia

In a Ti:Sa laser with an absorbing with Rb vapor cell stable self-starting passive mode-locking is demonstrated. We show that the mode-locking appears due to self-induced transparency (SIT) in the Rb cell, that is, the pulse in the Rb cell is a 2pi SIT pulse. For the best of our knowledge, in these experiments we present the first time demonstration of SIT mode-locking in laser systems, which was discussed only theoretical before.

## 1. INTRODUCTION

One of the most noticeable phenomena in nonlinear optics is the self-induced transparency (SIT) described theoretically and observed experimentally by McCall and Hann in [1]. In a medium with an atomic/molecular transition having the dipole moment $d_{12}$, and the relaxation times $T_1$ and $T_2$ of the population and polarization, short optical pulses at the resonance frequency can propagate without losses if the light-matte interaction is «coherent». That is, if the pulse duration is $\tau<T_2$, and, in addition, the pulse area S of the slow envelope $\varepsilon(t)$ defined as $S = \frac{d_{12}}{\hbar} \int \varepsilon(t) dt$ (here $\hbar$ is the Plank constant) satisfies the condition S=2π [2]. Such pulses are called 2π pulses. The mechanism of propagation without losses arises because the pulse transfers its energy to the atoms on the leading edge so that the atom are excited, after which the medium transfer the energy to the pulse back on the trailing edge, fully returning to the ground state; That is, a full Rabi oscillation takes place. This process takes place on the times less than relaxation times, therefore the energy is not dissipated and the energy losses are absent. There is another possibility to transmit the energy without losses on the resonant frequency. Namely, if the pulse area S=0 (so called 0π pulse), and the radiation takes form of two linked sub-pulses with the opposite signs of the envelope [2-4]. Spectrum of a 0π pulse consists of two peaks in the frequency space located on the both sides of resonance. Nevertheless, the term «SIT» is refereed typically to the 2π pulses [1,2].

SIT can be used to modulate nonlinear losses in a cavity and thus to achieve a passive mode-locking [5-13]. 0π or 2π pulses have the lowest losses in the absorber among all other pulse shapes, making them stable attractors of the system. In particular, if a small perturbation arises on the background of a 2π pulse is absorbed by the absorber and thus disappears. SIT-induced 2π pulses in cavity are more preferable than 0π ones because of more attractive temporal and spectral shape. Besides, as the pump power increases, the SIT induced pulse duration decreases. This is an extremely attractive property of SIT mode-locking, since there is no limit on the duration of Rabi oscillations, and thus there exists the possibility to generate few cycle pulses.

Despite of this, the possibility to use SIT for mode-locking has been discussed up to now only theoretically. One of the reasons is that, as it is typically believed, SIT mode-locking is rather difficult to realize practically since the experiments with SIT effect in pulse propagation are rather cumbersome [14]. The activity in this direction was limited also because the counter-intuitive nature of the fact that narrow lines may produce ultrashort pulses; in addition, most of the theoretical works were made in a two-level approximation which is not suitable for the ultrashort pulses, which also limited a practical interest.

Practically, to realize passive mode locking in two-section lasers (absorber and amplifier sections) resonant absorbers are long and successfully used [15-17] but, in contrast to a SIT absorber, work in the incoherent regime ($\tau > T_2$). For such saturable resonant absorbers the pulse duration is fundamentally limited by the width of the transition line in the absorber. Typically, the amplifier works also in the incoherent regime, which puts additional boundaries to the minimal pulse duration.

Up to now, any attempt to obtain mode-locking with SIT effects (2π pulses in absorber) and to show the absence of the usual pulse duration limits in the SIT case were unsuccessful. One should however mention the recent work [18] were a passive mode-locking in Ti:Sa laser with an intracavity Rb-87 cell was achieved using a semiconductor saturable absorbing mirror (SESAM). In that work the SIT pulses were indeed obtained in a Rb cell, however they were not the cause of mode-locking. In contrast, an opposite effect was observed: as the pump wavelength of the already mode-locked laser were tuned to the Rb resonance using an intacavity spectral filter, increasing of the pulse duration, decrease of the mode-locking stability and increasing of the repetition rate were observed in [18]. In our previous experiments, we observed mode-locking in a cw dye laser with a coherent absorber (cells with molecular iodine vapor) [19]. However, in those experiments, mode-locking occurred with formation of 0π zero-area pulses in absorber only, and not 2π SIT pulses, which could not be obtained in those experiments. Here we use an intracavity Rb cell as a nonlinear coherent resonant absorber in a Ti:Sa laser cavity. We show that the Rb cell causes a stable self-starting passive mode-locking at the low pump levels, much less than typically needed for the Kerr mode-locking. We claim that the mode-locking is caused by the SIT effect in the Rb cell, that is, the resulting mode-locked pulses are 2π pulses in the absorber. Although our pulses have picosecond duration, here we make, as we believe, is the first step towards realization of the mode-locking producing ultrashort pulses which are not limited by the absorber/amplifier line width.

## 2. EXPERIMENTAL SETUP

The setup we used is shown in Fig. 1. A Ti:Sa laser with several cavity designs was assembled on an optical table using standard components of a commercial laser («Technoscan»). To these components belonged a sapphire crystal (1) and spherical and plane mirrors (M1-M6). For tuning the wavelength of the laser a Lio filter, a Fabry-Perot etalon and prisms were used. The output of the pump laser VERDI V10 (Coherent) was focused, using the mirror (M) and the lens (L), through the mirror M1 into the Ti:Sa crystal (1). A chopper was put into the beam path, allowing to set the pump duration from 500 μs to CV to study the self-starting of the mode-locking. The Rb vapor cell with the natural mixture of isotopes and high quality windows oriented at the Brewster angle to the beam path to minimize the losses were used. The cell was placed before the output mirror of the cavity M5 and could be warmed and cooled. Mode-locking was observed in three different cavity designs, namely linear cavity (A), (B) and the ring cavity (C).

The wavelength tuning in the cavity design (A) was accomplished by rotating of the mirror M4, and in (B), (C) the Lio filter LF and the Fabry-Perot interferometer EFP were used for this goal. Laser intensity in dependence on time was registered using a high speed photodiode and an oscilloscope DSO 9104A (Agilent Technologies) with the temporal resolution 300 ps. The wavelength was determined by a spectrometer, the power was measured by the power meter «Maestro» (Standa). To estimate the pulse durations we used an interferometric autocorrelator based on the Michelson interferometer, as well as the Fabry-Perot interferometer with a variable base.

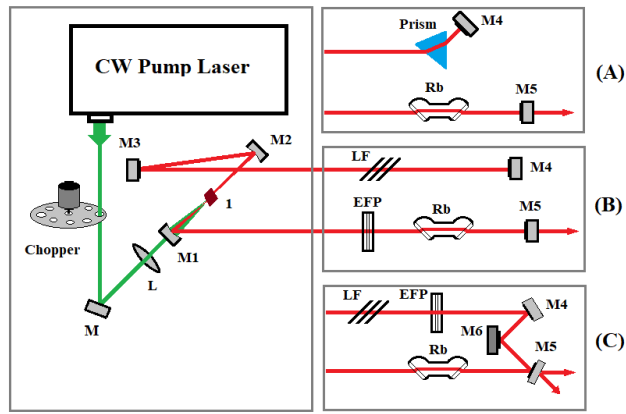

Fig. 1. Different cavity configurations of a Ti:Sa laser with a Rb vapor cell. (A), (B): Two variants of a linear cavity with different spectrally selective elements. (C): A ring cavity. Here 1 is a Ti:Sa crystal, M1,2 are the confocal mirrors of the cavity, M3 is an auxiliary mirror, M4,5 are the fully reflecting mirror and the output mirror (transparency around 8%), M6 is a fully reflecting auxiliary mirror, M is the mirror in the pump channel, L is a focusing lens, LF is the Lio filter, EFP is the Fabry-Perot etalon, Prism is a prism for tuning the lasing wavelength.

## 3. RESULTS

Mode-locking did not appear if the lasing wavelength was not tuned to the Rb transitions. Therefore, the intracavity spectral filters allowing to tune the lasing wavelength were used. Only with these spectral-selective elements, by tuning the wavelength to the resonant transitions DI (794 nm) and DII (780 nm) we were able to obtain mode-locking. In the linear cavity design (A) (see Fig. 1) for the wavelength tuning a prism was used, in (B) and (C) a Lio filter and a Fabry-Perot etalon were applied. Since no special measures to stabilize the laser were undertaken, mechanical instabilities and air flows lead, after several minutes of operation, to a drift of the filters from the cavity resonance; because of this, two filters were used to make the operation more stable. In this case, mode-locking was conserved for a quite extended time, from tens of minutes to an hour, which allowed to make measurements without tuning of the cavity (in the case (B)). The ring cavity (C) was more stable than (A) but less stable than (B). In the case (C), two counter-propagating waves were possible with the switching between them – a process which were difficult to control.

The mode-locking regimes were always accomplished by an intense luminescence of the Rb vapors in the cell. This luminescence were absent when no mode-locking were present. Besides, the mode-locking always disappeared if we cooled the cell with liquid nitrogen. In contrast, the mode-locking remained stable if the cell was warmed up to $70^0$ C. This allowed us to identify the Rb cell as the main ingredient needed for the mode-locking. Since the mode-locking was stably observed at the room temperature, the experiments described below were made without any warming of the cell.

The stable mode-locking was observed in all three cavity designs on the both Rb transitions DI and DII. In the ring cavity (C), the mode-locking was observed both if two counter propagating waves were present simultaneously and if only one of the waves were there. Examples of the oscillogrames of the output pulses in the mode-locking regime for the linear (B) and ring (C) cavity are shown in Fig. 2.

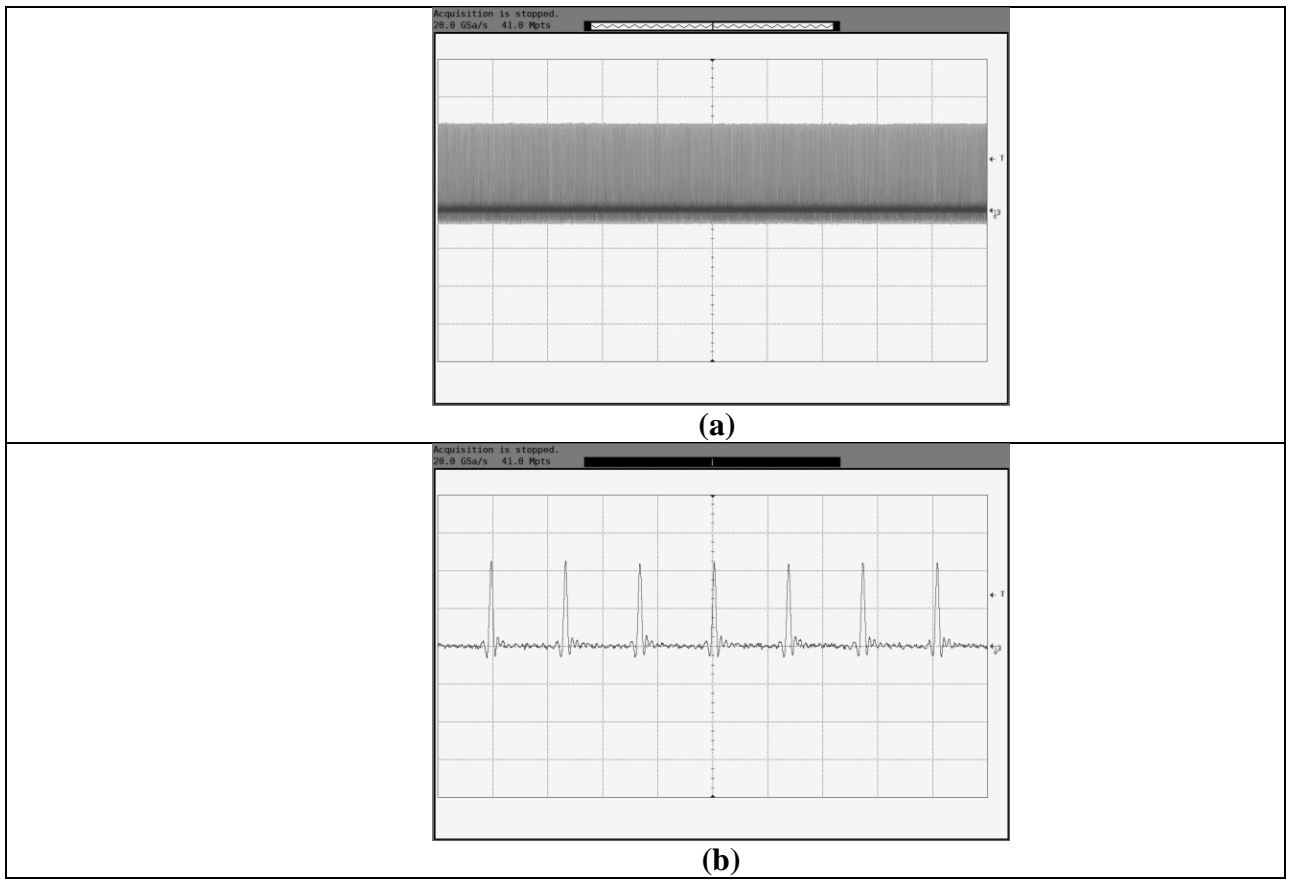

(a)

(b)

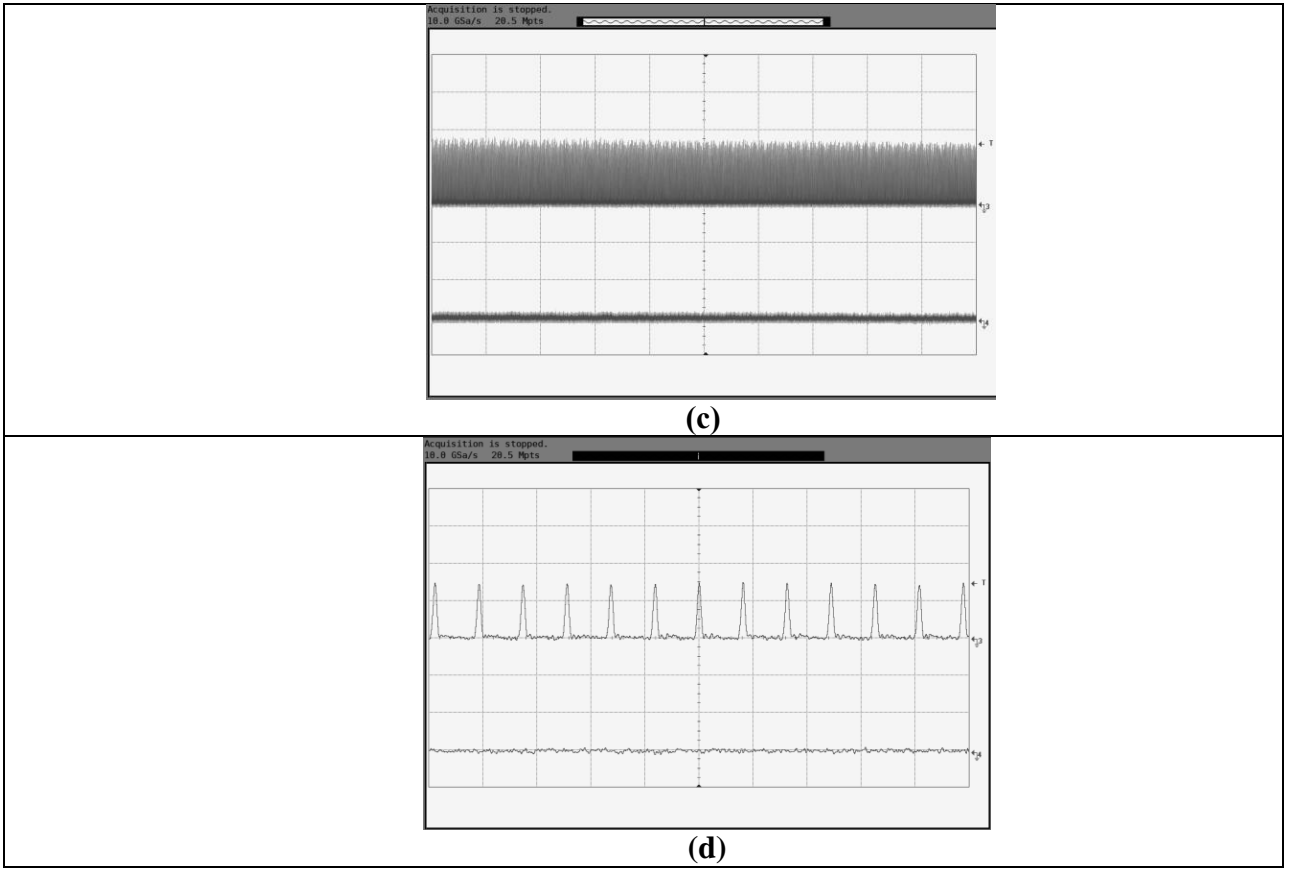

Fig. 2. Oscilloscope screenshots of the mode-locking regimes in a Ti:Sa laser with linear (a,b) and ring (c,d) cavity geometries, with the wavelength tuned to the transitions DII (a,b) and DI (c,d) of Rb with the resolution 200 μs/div in (a,c) and 5 μs/div (b,d). In (c,d), the upper (lower) line shows the clockwise- (anti-clockwise-) propagating wave.

It is also well known that Ti:Sa laser can act in a mode-locking regime induced by a so called Kerr lens [21,22]. In our experiments, the power was too low for such Kerr lens. In particular, the maximal output power in our experiments was not higher than 0.15 W (because of the highly transmitting output mirror (8.2 %) the intracavity power was also not large). We have observed that, without the Rb cell and selective elements, one needs much higher output power of 0.4 – 0.8 W to obtain the Kerr mode-locking. Importantly, in this later case the mode-locking could be only obtained using a mechanical disturbance (mechanically kicking the mirrors and/or the optical plate) and such mode-locking disappeared during several tens of μs. Therefore, we believe that the Kerr mechanism was not the reason of our mode-locking.

In the Rb vapors at the room temperature the decay rates are well known: $T_2 \approx 2T_1$, $T_1 \approx 27$ ns [20]. Repetition period (in our experiments from 4 ns to 7 ns) and observed pulse durations (in between 2 ns and 60 ps) are both less than $T_2$. Therefore, interaction of the radiation with the Rb transitions is clearly coherent. Because of this coherent character of the interaction, the losses in the Rb cell are zero for the pulses with the area $2\pi n$ with n=0,1,..: in this case the energy after the pulse passage is fully returned back to the radiation. The case of n=0 (zero area pulse) is excluded by the single-peak character of the pulses observed in experiments (since zero area pulses must have a double-peak structure as discussed before). On the other hand, it is known that the pulses with n>1 are unstable and split into several $2\pi$ ones [1,2]. Hence, we see that the most probable pulse configuration is a pulse with the area=$2\pi$ in the cell.

This fact can be furthermore checked experimentally. On the one hand, we can obtain the pulse duration from the intensity oscillograms or, when the duration is less than the resolution of the oscillograph, using an autocorrelator and a scanning Fabry-Perot interferometer. On the other hand, the duration of a $2\pi$ pulse can be estimated independently from the measured output power as the following: The Rabi period $T_{Rabi} = \frac{2\pi}{\Omega_{Rabi}}$ and Rabi frequency $\Omega_{Rabi} = \frac{d_{12}E}{\hbar}$ (here $E$ is the peak field strength; the values of $d_{12}$ for Rb are given in [18]) can be related to the output power $P_{out}$ via the expression: $E(ESU) = 27 \sqrt{W\left(\frac{W}{cm^2}\right)}/300$. The peak intensity is given by: $W = \frac{P_{out}T_{cav}}{T_r \tau \cdot \pi (D/2)^2}$, were $T_r$ is the transmittivity of the cavity mirror, $T_{cav}$ is the roundtrip time, $\tau$ is the pulse duration, $D$ is the diameter of the beam in the cell. If, for simplicity, we assume a rectangular pulse shape, for a $2\pi$ pulse we have $\tau = T_{Rabi}$. Combining the above expressions we will obtain:

$$\tau = \left(\frac{2\cdot\pi\cdot h}{d_{12}}\right)^2 \cdot \frac{T_r \cdot \pi \cdot \left(\frac{D}{2}\right)^2}{T_{cav}} \cdot \frac{1}{P_{out}} \qquad (1)$$

Eq. (1) relates the duration of $2\pi$ pulses with the parameters, all of which are accessible for our measurement with a good precision (except the pulse diameter which was not constantly monitored despite it could slightly vary with the pulse power). In Fig. 3 we show the pulse durations $\tau$ obtained using Eq. (1) for the line DI with the direct experimental measurements of $P_{out}$.

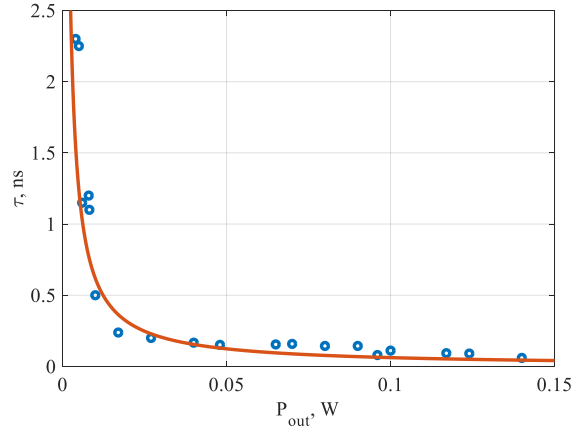

Fig. 3. The dependence of the mode-locked pulse duration for transition D1 in Rb on the output power. Points show the direct measurements whereas the line show the prediction of Eq. (1).

The results of Fig. 3 demonstrate a rather good agreement between the pulse durations expected for $2\pi$ pulses and measured in our experiment, which confirms the the SIT character of the pulses in the Rb cell. As seen, in this case the pulse duration is inversely proportional to the output power.

Another property of $2\pi$ pulses which can be experimentally checked is the independence of the pulse area on the peak intensity. In Fig. 4, we show the dependence of the pulse area obtained directly from the pulse shape given by the oscillograms on the peak voltage (representing the peak intensity). It can be seen that despite the peak intensity changes about one order of magnitude no significant change in the pulse area is observed.

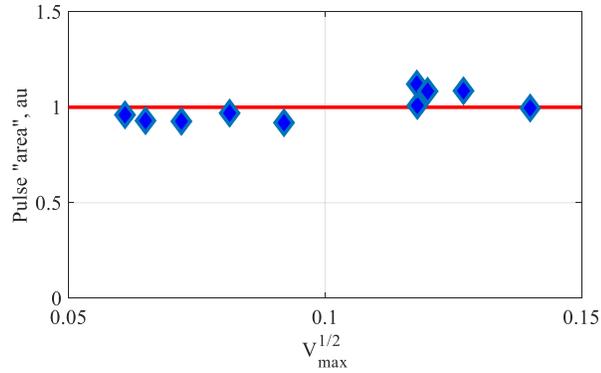

Fig.4. The dependence of the pulse area (in arbitrary units) on the square root of the peak voltage of the mode-locked pulse (which represents the peak field strength).

The important peculiarity of the SIT mode-locking is its possibility to self-start as was suggested by theoretical consideration [12]. Here we confirm the self-starting of our experimental scheme using a chopper which periodically broke the pump beam, allowing to study how the mode-locking regimes sets up. The chopper was installed on the path of the pump beam allowing to record the whole pulse train inside of every chopper cycle. We observed that the mode-locking is stable and reestablishes reliably in the beginning of every cycle. An example of a typical pulse train is given in Fig. 5. It is quite different from a passive saturable absorber mode-locking, where according to the fluctuation model [23] the more-locking develops from intensity fluctuations that arise at the initial stage of generation from broadband radiation. Saturable absorber plays the role of a "discriminator" [24] of weak short pulses, which, combined with amplification, isolates and forms a single fluctuation spike turning into a mode-locked pulse.

In a sharp contrast, the coherent absorber does not play a role of a discriminator. The scenario of the mode-locking emergence is complicated and quite different from the one for the standard passive mode-locking with saturable absorber (see Fig. 5).

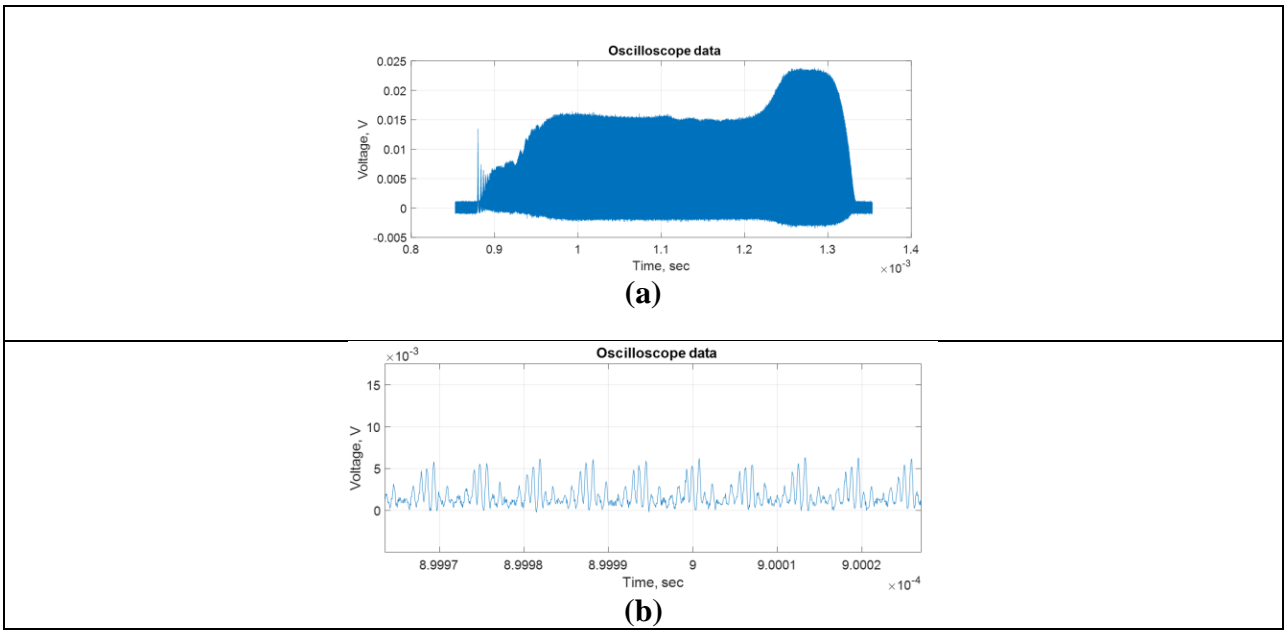

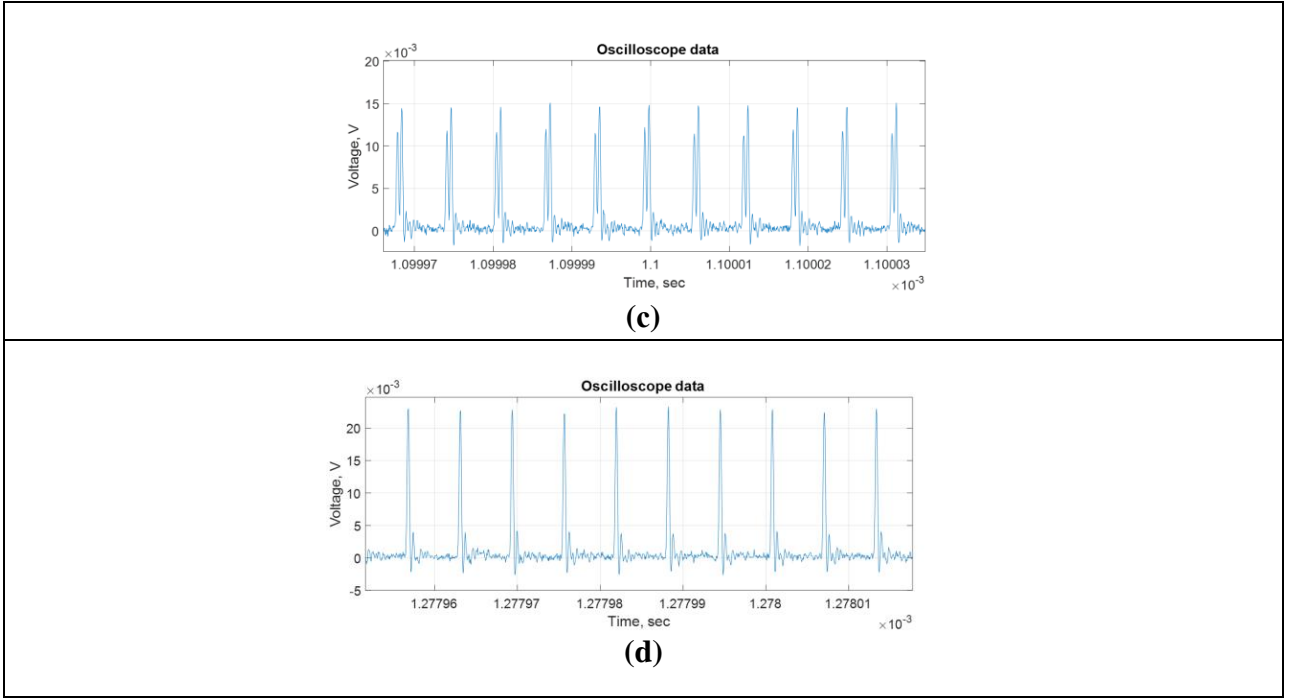

Fig. 5. Oscillograms of the mode-locked lasing during the time interval when the chopper opens the pump, showing reliable self-start of the mode-locking mode on the rubidium DII transition. (a) - Lasing in the whole interval of the open pump. (b) - (d) Transition from several zero-area pulses per roundrtip (b) to a single zero-are pulse (c), and then to a single $2\pi$ SIT pulse (d).

As it is seen in Fig. 5, at the initial stage, several low amplitude zero-area pulses are formed per roundtrip (Fig. 5b). During the further evolution, a single zero-area pulse survives (Fig. 5c), having also significantly higher amplitude. Zero-area pulses have a form of two coupled pulses with the spectrum containing a deep at the transition frequency. This deep was also observed in the interferograms (not shown). At the final stage (Fig. 5d), this pulse turns into a single $2\pi$ one. This $2\pi$ pulse finally remains stable up to the point when the chopper breaks the pump. Our oscillograms are able to distinguish clearly the main peak in the $2\pi$ pulse; The oscillations observed at the trailing edge of the pulses we attribute to transients in the photodetector circuit and oscilloscope. This set of dynamical transitions remain from one pulse train to another, although exact transitions time vary from train to train, and the whole transition to the stable takes some time. Nevertheless, the final state with a single $2\pi$ pulse per roundtrip is achieved at every chopper cycle. Thus, the transition to the mode-locking state shows a reach intriguing dynamics with the stable mode-locking as the final stage. The detailed study of the transient dynamics is interesting but beyond the scope of this article.

## 4. DISCUSSION

As it is known, experiments with self-induced transparency (SIT) in atomic media in a propagation geometry are typically relatively complex. Therefore, it was widely believed up to now that the SIT-based mode-locking should be even more complicated in realization. However, as we see here, it is quite easy to obtain very stable mode-locking using quite standard techniques. In our experiments we used a natural mixture of rubidium isotopes. Nevertheless, few points are to be cared about: The requirements to the optical quality of the cell windows are quite high. In our experiments we oriented them at the Brewster angle. Besides, rubidium vapor may settle on the windows creating thereby a thin metallic film invisible by eye. It creates losses and prevents lasing. Therefore, it is necessary to ensure that Rb is absent on the cell windows, or remove it by heating. Finally, the beam cross section in the cell must be as constant as possible, otherwise the pulse will not have the same area across the cell, the losses grow dramatically and it becomes impossible to obtain the mode-locking.

If these requirements are satisfied, it is not difficult at all to obtain the SIT mode-locking: A Rb cell at the room temperature is to be placed in the cavity of an arbitrary Ti:Sa laser near the output mirror, and the mirror must be adjusted to produce lasing. After that, the intracavity spectral filters are to be tuned to the transition of rubidium. Mode-locking is accompanied by heavy luminescence in the cell, which is easy to see with the help of an IR visualizer. Besides the regular mode-locking it is possible also to get chaotic spiking.

In our experiments, we placed the cell inside the cavity of a commercial Mira Optima 900-D (Coherent) laser, made readjustment and obtained mode-locking at rubidium transitions. The maximum output power was 1.2 W.

Note that the wavelength range of the Ti:Sa laser allows to reproduce coherent mode-locking not only in rubidium but also in cesium or potassium vapors.

## 5. CONCLUSIONS

We have demonstrated a passive self-induced transparency (SIT) mode-locking in a Ti:Sa laser, that is, mode-locking arising due to coherent interaction of radiation with resonant transitions in an intracavity rubidium vapor cell which in this case played a role of a coherent absorber. For the best of our knowledge, our results are the first experimental demonstration of SIT-based mode-locking regime. We have shown that the area S of the pulses, regardless of their power, is constant and is indeed equal to $2\pi$, that is, pulses of self-induced transparency in the Rb cell take place.

Furthermore, we have demonstrated that the mode-locking self-starts very stably, nevertheless demonstrating a reach dynamics in transition from non-lasing to the mode-locking state. The minimal duration of the mode-locked pulses was two orders of magnitude shorter than the time $T_2$ of the absorber. That is, the relaxation time $T_2$ of the absorber does not limit the pulse duration. Finally, our mode-locking was obtained at the input powers which are an order of magnitude less than needed for the Kerr mode-locking.

**Funding**. I.B. is thankful to the DFG (Deutsche Forschungsgemeinschaft) for the financial support (Project BA 4156/4-2) as well as Germany's Excellence Strategy within the Cluster of Excellence PhoenixD (EXC 2122, Project ID 390833453).
**Acknowledgment.** The Investigations were performed at the Center for Optical and Laser Materials Research. St. Petersburg State University.